\documentclass[12pt]{amsart}
\usepackage{amssymb}
\usepackage{latexsym}
\usepackage{epic, eepic, graphicx}
\usepackage{graphicx} \DeclareGraphicsExtensions{.ps}

\newcommand{\tA}{{\widetilde A}}
\newcommand{\tB}{{\widetilde B}}

\newcommand{\ep}{\epsilon}
\newcommand{\bep}{\boldsymbol{\epsilon}}

\newcommand{\CC}{{\mathbb C}}

\newcommand{\NN}{{\mathbb N}}

\newcommand{\RR}{{\mathbb R}}

\def\t2{{\mathbb T}^2}

\newcommand{\ZZ}{{\mathbb Z}}

\newcommand{\CIc}{{\mathcal C}^\infty_{\rm{c}} }

\newcommand{\cD}{{\mathcal D}}
\newcommand{\F}{{\mathcal F}}
\newcommand{\G}{{\mathcal G}}
\newcommand{\Hh}{{\mathcal H}}
\newcommand{\cA}{{\mathcal A}}
\newcommand{\cB}{{\mathcal B}}
\newcommand{\cC}{{\mathcal C}}

\newcommand{\Oo}{{\mathcal O}}

\newcommand{\cV}{{\mathcal V}}
\newcommand{\cW}{{\mathcal W}}

\newcommand{\vol}{\operatorname{vol}}

\newcommand{\half}{\frac{1}{2}}

\newcommand{\tr}{\operatorname{tr}}
\newcommand{\Res}{\operatorname{Res}}
\newcommand{\Spec}{\operatorname{Spec}}

\renewcommand{\Im}{\mathop{\rm Im}\nolimits}

\newcommand{\Op}{\operatorname{Op}}

\newcommand{\ra}{\rangle}
\newcommand{\la}{\langle}
\newcommand{\hn}{\Hh_{N}}
\newcommand{\bver}{\begin{verbatim}}

\newcommand{\defeq}{\stackrel{\rm{def}}{=}}
\def\hto0{\xrightarrow{h\to 0}}

\theoremstyle{plain}

\theoremstyle{definition}

\numberwithin{equation}{section}

\newcommand{\bequ}{\begin{equation}}

\def\bbbone{{\mathchoice {1\mskip-4mu {\rm{l}}} {1\mskip-4mu {\rm{l}}}
{ 1\mskip-4.5mu {\rm{l}}} { 1\mskip-5mu {\rm{l}}}}}

\newcommand{\set}[1]{\left\{\,#1\,\right\}}

\def\squarebox#1{\hbox to #1{\hfill\vbox to #1{\vfill}}}

\usepackage{epic, eepic}

\title
[Fractal Weyl laws in chaotic scattering]
{Fractal Weyl laws in discrete models of chaotic scattering}
\author[S. Nonnenmacher]
{St\'ephane Nonnenmacher}
\author[M. Zworski]
{Maciej Zworski}
\address{Service de Physique Th\'eorique, 
CEA/DSM/PhT, Unit\'e de recherche associ\'ee au CNRS,
CEA/Saclay,\\
91191 Gif-sur-Yvette, France}
\email{snonnenmacher@cea.fr}
\address{Mathematics Department, University of California \\
Evans Hall, Berkeley, CA 94720, USA}
\email{zworski@math.berkeley.edu}

\begin{document}

\begin{abstract}
We analyze simple models of quantum chaotic scattering, namely 
quantized open baker's maps. We numerically compute the density of quantum resonances
in the semiclassical r\'egime. This density satisfies a
fractal Weyl law, where the exponent is governed by the (fractal) dimension of
the set of trapped trajectories. This type of behaviour is also expected in the 
(physically more relevant) case of Hamiltonian chaotic scattering. Within a 
simplified model, we are able to rigorously prove this Weyl law, and compute
quantities related to the ``coherent transport'' through the system, namely the conductance 
and ``shot noise''. The latter is close to the prediction of random matrix theory.
\end{abstract}

\maketitle

\section{Introduction}
\label{intr}

The study of resonances, or quasibound states, has a long tradition 
in theoretical, numerical, and experimental 
chaotic scattering -- see for instance \cite{G} 
and references given there. 
In this paper we discuss the laws for the density of resonances at
high energies, or in the semiclassical limit, and the closely related
asymptotics of conductance, Fano factors, and ``shot noise''.
Our models are based on a quantization of open 
baker maps \cite{BaVo,Sa,SaVa} and we focus on 
{\em fractal Weyl laws} for the density of resonances. These laws
have origins in the mathematical work on counting resonances \cite{ZwN}.

In \S\ref{obq} we present the 
compact phase space models for chaotic scattering (open baker's maps)
and their discrete quantizations.
The numerical results on counting of quantum resonances showing 
an agreement with fractal Weyl laws are given in \S\ref{nr}. In 
\S\ref{cm} we discuss a model which is simpler on the quantum level
but more complicated on the classical level (this model can also be interpreted
as an alternative quantization of the original baker's map, see \S\ref{dr}). 
In that case we 
can describe the distribution of resonances very precisely 
(\S\ref{dr}), showing perfect agreement with the fractal Weyl law.
We also find asymptotic expressions for the conductance and
the Fano factor (or the ``shot noise'' factor). The fractal 
Weyl law appears naturally in these asymptotics and an interesting
comparison with the random matrix theory is also made (\S\ref{ff}). 

To put the fractal Weyl law in perspective we 
review the usual Weyl law for the density of states in the 
semiclassical limit. Let $ H ( q, p ) = p^2 + V ( q ) $ be a
Hamiltonian with a confining potential $ V $ and let $ E $ be a
nondegenerate energy level,  
\begin{equation}
\label{eq:nod}
  H ( q , p ) = E \ \Longrightarrow \ \nabla H ( q , p ) \neq 0 \,.
\end{equation}
Assume further that the union of periodic orbits of the Hamilton flow on the
surface
$ H^{-1}(E) $ has measure zero with respect to the Liouville measure.
Then the spectrum of the quantized Hamiltonian,
\begin{equation}
\label{eq:Hamilton}
 \widehat H = -\hbar^2 \Delta + V ( q ) \,, \quad  q \in \RR^n \,, 
\end{equation}
near $ E $ satisfies,
\begin{equation}
\label{eq:weyl-closed}
 \# \set{\Spec ( \widehat H )  \cap [E- \rho \hbar, E+ \rho \hbar]}  = 
\frac{2 \rho  \hbar }
{ ( 2 \pi \hbar )^n } \int \delta\big(H(q,p)-E\big)\  d q \, dp +
o ( \hbar^{-n + 1 } ) \, ,
\end{equation}
see \cite{Ivr} for references to the mathematical literature on this 
subject. 

Suppose now that $ V ( q ) $ is {\em not} confining. The most 
extreme case is given by $ V(q) $ vanishing outside a compact
set. An example of such a potential with $ q \in \RR^2 $ is given in 
Fig.~\ref{f:3b}. In that case the eigenvalues are replaced by 
{\em quantum resonances} which can be defined as the poles of the 
meromorphic continuation of Green's function, $  G( z ; q', q )  $,
from $ \Im z > 0 $ to $ \Im z \leq 0 $. By Green's function we mean
the integral kernel of the resolvent:
\bequ\label{e:resolvent}
( z - \widehat H )^{-1} u ( q' ) = \int_{\RR^n } G( z ; q', q )\, u ( q )\, d q \,, 
\quad u \in \CIc ( \RR^n ) \,.
\end{equation}
We denote the set of resonances by $ \Res ( \widehat H ) $.
Near a nondegenerate energy level \eqref{eq:nod} we have the following 
bound (compare with \eqref{eq:weyl-closed} for a closed system) \cite{Bo}:
\begin{equation}
\label{eq:Weyl-improved} 
\#\set{ \Res ( \widehat H) \cap \big([ E - \rho \hbar , E + \rho \hbar ] - i[ 0 ,  \gamma \hbar ]\big) } 
 \leq C ( \rho , \gamma )\, \hbar^{-n+1} 
\,.\end{equation}
When the interaction region is separated from infinity by a barrier,
this bound is optimal since resonances are well approximated by 
eigenvalues of a closed system.
In that case the classical trapped set,
\begin{equation}
\label{eq:trap}
 K_E \defeq \{ ( q, p )\in H^{-1}(E) \; : \; \Phi_H^t ( q , p ) \not \rightarrow 
\infty \,, \quad  t \rightarrow \pm \infty \} \,, 
\end{equation}
has a non-empty interior in $ H^{-1}(E) $, so that its dimension
is equal to $ 2 n - 1 $.

Suppose now that the classical flow of the Hamiltonian $ H$ is
hyperbolic on  $ K_{E} $, as is the case for instance in some energy range for 
the 2-D potential represented in Fig.~\ref{f:3b} \cite{Mor,SjDuke}.
%%%%%%%%%%%%%%%%%%%%%%%%%%%%%%%%%%%%
\begin{figure}[htbp]
\begin{center}
\rotatebox{-90}{\includegraphics[width=3in]{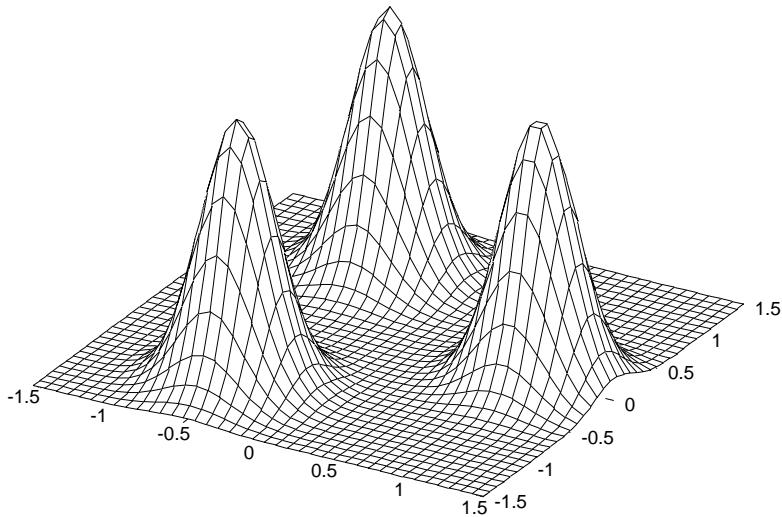}}
\end{center}
\caption
{\label{Fblock} A three bump potential exhibiting hyperbolic dynamics
on a certain energy range.}
\label{f:3b}
\end{figure}
%%%%%%%%%%%%%%%%%%%%%%%%%%%%%%%%%%%%
Following the original work of Sj\"ostrand \cite{SjDuke}, 
the general upper bound \eqref{eq:Weyl-improved} is replaced by a bound involving
the {\em upper Minkowski dimension} of $ K_E $:
\[
\dim K_E  = 2n -1 - \sup \set{ c \; : \; \limsup_{ \epsilon \rightarrow 0 }
\epsilon^{-c} \vol ( \{ \rho \in \Sigma_{E} \; : \; d ( K_E , \rho )
< \epsilon \} ) < \infty } \,.
\]
We say that $ K_E $ is of {\em pure dimension} if the supremum is
attained. For simplicity we assume that this is the case. Then
under the assumption of hyperbolicity of the flow, we have \cite{SjZw04}
\begin{equation}
\label{eq:fractal-hamil} 
\# \set{\Res ( \widehat H) \cap \big([ E - \rho \hbar , E + \rho \hbar ] - i[ 0 ,  \gamma \hbar ]\big) } 
\leq C ( \rho , \gamma )\, \hbar^{-\mu_E} 
\,, \quad   2 \mu_E + 1 = \dim K_E \,.
\end{equation}
This bound is expected to be optimal even though it is not clear
what notion of dimension should be used for the lower bounds. 
The best chance lies in cases in which $ K_E $ has a particularly 
nice structure. A class of Hamiltonians for which that happens is
given by quotients of hyperbolic space by convex co-compact discrete
groups \cite{GLZ}.

\renewcommand\thefootnote{\dag}%

A fractal Weyl law for the density of 
resonances in larger regions are easier to verify and more likely 
to hold in general:
\begin{equation}
\label{eq:fractal-large}
\# \set{\Res ( \widehat H) \cap \big([ E - \delta , E + \delta ] - i[ 0 ,  \gamma \hbar ]\big) }
\sim C( \delta , \gamma)\, \hbar^{-\mu_E - 1 } \,, \qquad \delta>0\ \text{fixed}.
\end{equation}
The precise meaning of $ \sim $ is left vague in this 
conjectural statement. The exponent in this relation has been 
investigated numerically in a variety of settings and the results
are encouraging \cite{L}.

\subsection*{Acknowledgments}
The first author thanks Marcos Saraceno for his insights and comments on the 
various types of quantum bakers. He is also grateful to UC Berkeley for the hospitality 
in April 2004. Generous support of both authors by the National Science
Foundation under the grant DMS-0200732 is also gratefully acknowledged.

%%%%%%%%%%%%%%%%%%%%%%%%%%%%%%%%%%%%
%%%%%%%%%%%%%%%%%%%%%%%%%%%%%%%%%%%%
\section{Open baker maps and their quantizations}
\label{obq}

We consider $ \t2 = [0,1) \times [0,1) $,
the two-torus, as our classical phase
space. Classical observables are functions on $ \t2 $ and the
classical dynamics is given in terms of an ``open symplectic map'' $B$, that is
a map defined on a subset $\cD\subset \t2$, which is invertible and 
canonical (area and orientation preserving) from $\cD$ to $B(\cD)$. The points of
$\t2\setminus \cD$ are interpreted as ``falling in the hole'', or
``escaping to infinity''

Following a construction performed in \cite{SaVa}, 
we will be concerned with open versions of the baker's map, obtained by
restricting the ``closed'' baker's map to a subdomain of $\t2$, union of
vertical strips. As an example, 
if we restrict the 3-baker's map $ A_3 $:
\begin{gather}
\label{eq:honest}
\begin{gathered}
A_3( q, p ) \defeq (q',p') =
 \left\{  \begin{array}{lll}
q' = 3 q\,, & p' = p/3\,, & \text{if}\  0 \leq q < 1/3 \\
q' = 3 q -1 \,, & p' = (p+1) /3\,, & \text{if}\  1/3 \leq q < 2/3 \\
q' = 3 q - 2\,, & p' = (p+2) /3\,, & \text{if}\  2/3 \leq q < 1  
\end{array}\right. \,.
\end{gathered}
\end{gather}
to the domain $\cD_3=\t2\setminus\{1/3\leq q<2/3\}$, we obtain the open 3-baker's map $B_3$:
\begin{equation}
\label{eq:exb}  
\forall (q,p)\in \cD_3,\quad B_3(q,p)=(q',p') =
\left\{ \begin{array}{lll}
q' = 3 q\,, & p' = p/3\,, & \text{if}\ \ 0 \leq q < 1/3 \\
q' = 3 q - 2\,, & p' = (p+2) /3\,, & \text{if}\ \ 2/3 \leq q < 1 \,.  \end{array}
\right. 
\end{equation}
This open map admits an inverse $B_3^{-1}$, which is a canonical map from $B_3(\cD_3)$ to $\cD_3$.
In this paper 
we will present numerical results for an open 5-baker's map,
defined as
\begin{equation}
\label{eq:exA}  
B_5(q,p)=(q',p') \defeq 
\left\{ \begin{array}{lll}
q' = 5 q-1\,, & p' = (p+1)/5\,, & \text{if}\ \ 1/5 \leq q < 2/5 \\
q' = 5 q - 3\,, & p' = (p+3) /5\,, & \text{if}\ \ 3/5 \leq q < 4/5 \,.  
\end{array}\right. 
\end{equation}

One can think of $A_3 $ as model for a Poincar\'e map for a
2-D closed Hamiltonian system. Removing the domain
$ \{1/3 \leq q < 2/3\} $ from the torus corresponds to opening the
system: the points in this domain will escape through the hole, that is, never
come back to the Poincar\'e section. In the context of mesoscopic quantum dots,
such an opening is performed by connecting a {\it lead} to the dot, through which
electrons are able to escape (see \S\ref{ff}).

For open maps such as $B=B_3$, we can define the 
{\em incoming and outgoing tails}, made of points
which never escape in the forward (resp. backward) evolution:
\[ \begin{split}
 x \in \Gamma_- \subset\t2 \ &\Longleftrightarrow \ \forall n\geq 0,\ B^n(x)\in \cD\,.\\
x\in\Gamma_+ \ & \Longleftrightarrow \  \forall n\geq 0,\ B^{-n}(x)\in B(\cD)\,.
\end{split} \]
In the case of the map \eqref{eq:exb}, $ \Gamma_- = \cC_3 \times [0,1) $,
$ \Gamma_+ = [0,1) \times \cC_3 $, where $ \cC_3 $ is the standard $ \frac13$-Cantor
set on the interval (see Fig.~\ref{f:K}).

%%%%%%%%%%%%%%%%%%%%%%%%%%%%%%%%%%%%
\begin{figure}[htbp]
\begin{center}
%{\includegraphics[height=3.75cm]{gm3.eps}\quad
%    \includegraphics[height=3.75cm]{gp3.eps}\quad
%    \includegraphics[height=3.75cm]{K.eps}}
\includegraphics[width=16cm]{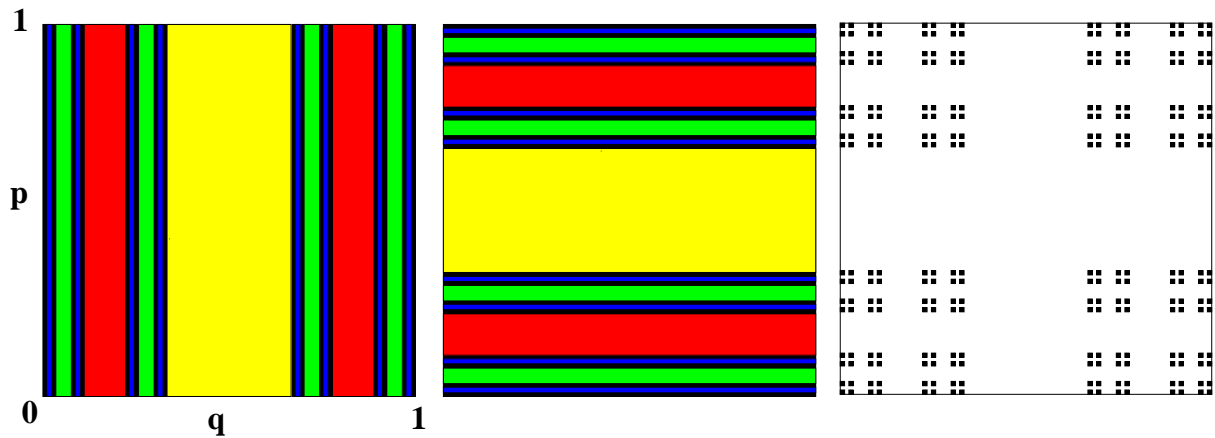}
    \caption{\label{f:K} We show, from left to right, approximations of the in/outcoming
tails $ \Gamma_- $, $ \Gamma_+ $ and the trapped set $ K $ for the open 3-baker $B_3$. 
On the left and central
plots, each color corresponds to points escaping at the same time.}
\end{center}
\end{figure}
%%%%%%%%%%%%%%%%%%%%%%%%%%%%%%%%%%%%

In analogy with \eqref{eq:trap},
we also define the {\em trapped set} $ K = \Gamma_+ \cap \Gamma_- $
and, for any point $ x\in K $, its stable and unstable manifolds $ W_\pm(x) $.
In the case of the open 3-baker $B_3$, we easily check that
\[
\mu\defeq \dim \Gamma_- \cap W_+ = \dim \Gamma_+ \cap W_- =\half \dim K = 
\frac{\log 2}{ \log 3} \,.
\]

The quantization of the open map \eqref{eq:exb} is based on the quantization of
the ``closed'' baker's map $A_3$. That, in an outline, is
done as follows \cite{BaVo,Sa}. To any $N\in\NN$ we associate a space $\hn\simeq\CC^N$ of quantum
states on the torus. The components $\psi_j$, $j\in\ZZ_N=\set{0,\ldots, N-1}$ 
of a state $\psi\in\hn$ are the amplitudes
of $\psi$ at the positions $q=q_j=(j+\half)/{N}$, and we will sometimes use 
Dirac's notation $\psi_j=\la q_j|\psi\ra$. The choice of these ``half-integers positions'' is justified by
the parity symmetry $q\to 1-q$ they satisfy \cite{Sa}, and is further explained in \S\ref{nr}.
The scalar product on $\hn$ is the standard one on $\CC^N$:
\begin{equation}
\label{eq:inn}\forall \psi , \phi \in \hn ,\quad  \langle \phi | \psi \ra = 
\sum_{j=0}^{N-1}\overline \phi_j\, \psi_j \,.
\end{equation}
A classical observable depending on $ q \in \RR/ \ZZ  $ only, 
$ f = f ( q ) $, is obviously quantized as the multiplication operator
\[ 
\forall \psi\in\hn,\quad [\Op_N ( f ) \psi]_j = f \left( \frac{j+1/2}{N} 
\right)\, \psi_j\,.
\]
The {\em discrete Fourier transform} 
\bequ\label{e:Fourier1}
(\G_N)_{j,j'}=N^{-1/2}\,e^{-2 i \pi  (j+\half)( j'+\half)/N}\,,\quad j,j'=0,\ldots,N-1\,
\end{equation}
transforms a ``position vector'' $\psi_{j}=\la q_{j}|\psi\ra$ into the corresponding 
``momentum vector''
$\la p_j|\psi\ra=( \G_N \psi )_j$. The momenta are also quantized to
values $p_j=(j+\half)/{N}$, $j=0,\ldots,N-1$. 
Comparing the definition \eqref{e:Fourier1} with the (standard) Fourier transform 
on $ \RR $,
\[ 
\F_\hbar u ( p ) = \frac{1}{ \sqrt{ 2 \pi \hbar } } \int_\RR e^{ - i p q / \hbar }
u ( q ) dq \,,
\]
we see that the effective Planck's constant in the discrete model 
is $ \hbar = (2\pi N)^{-1} $. 

As a result, any observable 
$ g = g ( p )$ can be quantized as
$$
\Op_N ( g ) \psi = \G_N^*\, \text{diag}\Big(g \big(  (j+1/2)/{N} \big)\Big)\,\G_N\, \psi\,.
$$
The Weyl quantization on the torus generalizes this map
$ f \mapsto \Op_N ( f ) $ to any classical 
observable $ f $, that is 
any (smooth) function on the torus, 
in such a way that a real observable $ f $ is associated with 
self-adjoint operators, and 
\[ 
\frac{i}{\hbar} [ \Op_N ( f ) , \Op_N ( g ) ] = \Op_N ( \{ f , g \} ) + 
{\mathcal O} ( \hbar^{2} ) \,.
\]
Let us now consider the following family of unitary operators on
$\hn$, where $N$ is taken as a multiple of $3$:
\begin{equation}
\label{eq:UN}
\widehat A_{3,{\rm pos}}= A_{3,N} \defeq \G_N^* \left( \begin{array}{lll} \G_{N/3} & \ 0 & \ 0 \\
\ 0 &  \G_{N/3}  &  \ 0 \\
\ 0 & \ 0 & \G_{N/3} \end{array} \right) \,.
\end{equation}
Since $ \G_N $ exchanges position and momentum, the 
mixed momentum-position representation of $\widehat A_3$ is given by the matrix
\[ 
\widehat A_{3,{\rm mom-pos}}=\G_N\, A_{3,N}=\left( \begin{array}{ccc} \G_{N/3} & \ 0 & \ 0 \\
\ 0 &  \G_{N/3}  &  \ 0 \\
\ 0 & \ 0 & \G_{N/3} \end{array} \right)  \,.
\]
In terms of the quantized positions $ q_j $ and momenta $ p_k $,
the entries of this matrix are given by
\begin{gather*}
\big(\widehat A_{3,{\rm mom-pos}}\big)_{k\,j}=
\la p_k|\widehat A_3|q_j\ra=  \frac{1}{ \sqrt{ 2 \pi \hbar } } \
\exp \left(- \frac{i}{\hbar} ( 3 q_j - \ell ) \big( p_k - \frac{\ell}3 \big) \right) \,, \\ 
\frac{\ell}3 \leq  q_j < \frac{\ell+1}3\,, \quad  
\frac{\ell}3 \leq  p_k < \frac{\ell+1}3 \,, \quad \ell=0,1,2 \,,
\end{gather*}
and zero otherwise. One can then
observe \cite{BaVo} that for $\ell=0,1,2$, 
the function $S_\ell ( p' , q ) = ( 3 q - \ell ) ( p' - \ell/3 ) $
generates the canonical map $ ( q , p ) \mapsto ( q' , p') = ( 3q - \ell , p/3 + \ell/3 ) $ on the
domain $\{q,p'\in [\ell/3, (\ell+1)/3)\}$, that is, the map $A_3$ \eqref{eq:honest} on this domain. 
The matrix elements $\la p_k|\widehat A_3|q_j\ra$ therefore
exactly correspond to the Van Vleck semiclassical formula
associated with the map $A_3$. For this reason (and the unitarity of $\widehat A_3$),  
the operator $\widehat A_3$ was considered
a good quantization of $A_3$ by Balazs and Voros. 
A more precise description of the correspondence between $A_3$ and $\widehat A_3$, 
including the role played by 
the discontinuities of $A_3$, is explained in \cite[\S 4.4]{NZ}.

To quantize the open baker $B_3$ \eqref{eq:exb}, we truncate
the unitary operator $ \widehat A_3$ using the quantum projector on the
domain $\cD$, $ \Pi_{\cD} \defeq \Op_N (\bbbone_{\cD} ) $ \cite{SaVa}: in the position basis, we get 
\begin{equation}
\label{eq:BN} 
\widehat B_{3,{\rm pos}}= B_{3,N}\defeq A_{3,N}\, \Pi_{\cD} = 
 \G_N^* \left( \begin{array}{ccc} \G_{N/3} &  0 & \ 0 \\
\ 0 &  0  &  \ 0 \\
\ 0 &  0 & \G_{N/3} \end{array} \right) \,, \quad N\in 3\NN \,.
\end{equation}
This subunitary operator is a model for the quantization of a Poincar\'e
map of an open chaotic system \cite{SaVa}. The semiclassical r\'egime corresponds
to the limit $N\to\infty$. 
Similarly, the quantum open map associated with the
5-baker $B_5$ \eqref{eq:exA} is given by the sequence of matrices:
\begin{equation}
\label{eq:AN} 
B_{5,N} \defeq
 \G_N^* \left( \begin{array}{ccccc} 
0 &  0  &  0 & 0 & 0 \\
0 &\G_{N/5} &  0 &  0&  0 \\
0 &  0  &  0 & 0 & 0 \\
0 &  0 & 0 & \G_{N/5} & 0\\
0 &  0  &  0 & 0 & 0 
\end{array} \right) \,,\quad N\in 5\NN \,.
\end{equation}

Let us now describe the correspondence between the resonances
of a Schr\"odinger operator $\widehat H$, and the eigenvalues of our subunitary
open quantum maps $B_{3,N}$ or $B_{5,N}$ (denoted generically by $B_N$).

Since $B_N$ is obtained by truncating the unitary propagator $A_N$, it is
natural to consider the family of truncated Schr\"odinger propagators 
$\chi\, e^{- i t\widehat H /\hbar}\, \chi$, where $\chi(q)$ is a cutoff function 
on some compact set supporting the scatterer. Although the precise eigenvalues of these
propagators depend nontrivially on both $\chi$ and the time $t$, these propagators admit a
long-time expansion in terms of the resonances $z_j$ of $\widehat H$  \cite{BZ}.
At an informal level, one may write
$$
\chi\, e^{- i t\widehat H /\hbar}\, \chi \sim \sum_{z_j\in\Res(\widehat H)}\,
e^{- i t z_j / \hbar}\,\widehat R_j\,.
$$
On the other hand, the iterated open quantum map $(B_N)^n$ 
can obviously be expanded in terms of the eigenvalues $\lambda_j$ of $B_N$.
For this reason, it makes sense to model the
exponentials $e^{- i z_j / \hbar}$ by the eigenvalues $ \lambda_j$ of our open quantum map 
$B_N$. 

Upon this identification, the boxes in which 
we count resonances in \eqref{eq:fractal-hamil}, 
$  [ E - \rho \hbar , E + \rho \hbar ] - i[ 0 ,  \gamma \hbar ] $,  correspond
to the regions
\bequ\label{e:sector}
\cA_{r,\vartheta,\rho}\defeq 
\set{ 1\geq |\lambda | \geq r\,, \ | \arg ( \lambda \,e^{i \vartheta}) | \leq \rho }\,,
\qquad r=\exp ( - \gamma )\in(0,1)\,.
\end{equation}
These analogies induce a
conjectural fractal Weyl law for the quantum open bakers (\ref{eq:exb},\ref{eq:exA}) which we now
describe.

First of all, we consider the partial dimension of the trapped set of the
open map $B$:
$$
\mu = \frac{\dim K}{2} = \dim ( \Gamma_- \cap W_{+} ) \,.
$$
Then, for any $r\in(0,1)$, there should exist $C(r)\geq 0$ (a priori, depending on the map $B$)
such that, in the semiclassical limit, 
the number of eigenvalues of $B_N$ in the sectors 
\eqref{e:sector} behaves as
\begin{equation}
\label{eq:weyl-1} 
\# \set{ \lambda \in \Spec ( B_N )\cap \cA_{r,\vartheta,\rho}}  \simeq  \ \frac{\rho}{2\pi}\, C(r) \,
N^{\mu }  \,\qquad N\to\infty\,. 
\end{equation}
The angular dependence ${\rho}/({2\pi})$ on the RHS means that the distribution
of eigenvalues is expected to be asymptotically angular-symmetric.

In \cite{SaVo}, the quantum 2-baker $A_{2,N}$ was decomposed
into the block $\G_N^*\begin{pmatrix}\G_{N/2}&0\\0&0\end{pmatrix}$ and the complementary one. 
The spectral determinant for the unitary map, $\det(1-z A_{2,N})$, was then expanded 
in terms of these blocks. 
Although the classical open map associated with each block is quite simple (all points except a fixed one
eventually escape), the spectrum of each block was found to be rather complex, and quite different from
semiclassical predictions.

A scaling of the type \eqref{eq:weyl-1}
was conjectured in \cite{schomerus} for another chaotic
map, namely the open kicked rotator. This conjecture
was then tested numerically, and a good agreement was observed. The scaling law $N^\mu$ was explained 
heuristically by counting the number of quantum states
in an $\hbar$-neighbourhood of the incoming tail $\Gamma_-$ (so that $\mu$ was
effectively the dimension of $\Gamma_-$). For the kicked rotator, the fractal exponent $\mu$ 
was not known analytically, and the authors related it 
to the \emph{mean dwell time} of the dynamics, that is, the average time
spent in the cavity before leaving it: $\mu\approx 1-(\lambda \tau_{\rm dwell})^{-1}$ ($\lambda$ is
the mean Lyapounov exponent).
For our open baker's maps $B_3$, the dwell time is $\tau_{\rm dwell}=3$, so the above formula
is not valid: $\log 2/\log 3\neq 1-1/(3\log 3)$.
However, the above formula should give a good approximation of $\mu$ 
for a system with a large dwell time (that is, a small opening). 

In \S\ref{nr} we provide numerical evidence for the validity of \eqref{eq:weyl-1}
in the case of the open 5-baker $B_5$ \eqref{eq:exA}, 
at least when taking $N$ along {\em geometric subsequences}.
In \S\ref{dr} we then construct a related quantum
model, for which we can prove this Weyl law and calculate $C ( r )$ explicitly for 
inverse Planck's constants of the form $N=5^k$, $k\in\NN$.

%%%%%%%%%%%%%%%%%%%%%%%%%%%%%%%%%%%%%%%%%%%%%%%
%%%%%%%%%%%%%%%%%%%%%%%%%%%%%%%%%%%%%%%%%%%%%%%
\section{Numerical results}
\label{nr}

We numerically computed the spectra 
of several open baker's maps; in \cite[\S 5]{NZ} we showed the numerical results
concerning the $3$-baker $B_3$ \eqref{eq:exb}. For a change, we will
discuss here the open $5$-baker \eqref{eq:exA}, quantized in \eqref{eq:AN}.
For this open map, the partial dimension of the repeller is
$\mu= \log 2/{\log 5}=0.4306765...$ Compared to the 3-baker, this smaller exponent
implies that the spectrum of $\widehat B_5$ is expected to be much sparser than that of
$\widehat B_3$. For this reason, we will represent the spectra using a logarithmic
scale (see Fig.~\ref{f:spect3be}), and consider regions $\cA_{r,\vartheta,\rho}$ 
for values of $r$ ranging from $r=0.5$ down to about $r=0.001$.

Let us now briefly explain the choice of ``half-integer quantization'' for the quantum
positions and momenta \cite{Sa}. The open map $B_5$ 
is symmetric with respect to the
parity transformation $\Pi(q,p)=(1-q,1-p)$: $\Pi\circ B_5= B_5\circ \Pi$. 
The choice of quantization is made so that
the associated quantum map $\widehat B_5$ also possess this symmetry, that is, it commutes with the 
quantum parity operator $\widehat\Pi$ defined as $\widehat\Pi|q_j\ra=-|1-q_{j}\ra=-|q_{N-1-j}\ra$. 
We can then separately diagonalize the even and odd parts 
$$
\widehat B_{5,{\rm{ev}}}=\widehat B_5\circ ({1+\widehat\Pi})/{2} \ \text{ and } \ 
\widehat B_{5,{\rm{odd}}}=\widehat B_5\circ({1-\widehat\Pi})/{2} \,. 
$$
Both these operators have rank $N/5$: together, 
they give the full nontrivial spectrum of $\widehat B_5$. 
We checked that the odd spectrum has the  same characteristics as the even one, 
so we only describe the properties of the latter. It is expected to satisfy
the following fractal law (consequence of \eqref{eq:weyl-1}):
\bequ\label{e:fractal}
n(N,r)\defeq\#\{\Spec(B_{5,N,{\rm ev}})\cap \cA_r\}\simeq 
C(r)\,(N/5)^{\log 2/\log 5}\,, \quad \cA _r =\{|\lambda|>r\}\,,\quad N\to\infty\,.
\end{equation}
The simplest set of $N$'s to test this fractal Weyl law
is given by geometric sequences of the type $ \{ N_o\times 5^k,\ k=0,1,\ldots\} $: the law 
\eqref{e:fractal} means that the number of eigenvalues 
{\em doubles} when $k\to k+1$. 
In table~\ref{table2} we give some of the numbers $n(N,r)$ along
the sequence $N\in \{5^k\times 20\}$, for some
selected values of $r$.
%%%%%%%%%%%%%%%%%%%%%%%%%%%%%%%%%
\begin{table}[htbp]
\begin{center}
\begin{tabular}{||l|l|l|l|l|l|l||}
\hline
\  &  \ &   \  &    \  &    \  &    \  &   \ \\
$ N = 20\times 5^k $ & $ r = 0.5$ & $ r = 0.1$  & $ r = 0.05 $ & $ r = 0.01 $  &
$ r = 0.05 $ & $ r = 0.001$  \\
\  &  \ &   \  &    \  &    \  &    \  &   \   \\
\hline
$ k=0 $  &  4 & 10 & 10 & 13 & 14 & 16  \\
\hline
$ k = 1 $ & 7 & 19 & 19 & 25 & 27 & 35 \\
\hline
$ k = 2 $ & 15 & 36 & 36 & 48 & 55 & 122  \\
\hline
$ k = 3 $ & 30 & 69 & 69 & 104 & 216 & 402 \\
\hline
\end{tabular}

\bigskip

\caption{Number of even-parity eigenvalues of $ B_{5,N}$ in
$ \{| \lambda| > r\} $, for $ N $ along the sequence $\{20\times 5^k\}$.}
\label{table2}
\end{center}
\end{table}
%%%%%%%%%%%%%%%%%%%%%%%%%%%%%%%%%
Along each column with $r\geq 0.01$, the numbers approximately double at each step $k\to k+1$, which
seems to confirm the law \eqref{e:fractal}. The fact that this law fails for the 
small radii $r=0.005,\,0.001$ may be explained as follows: according to \eqref{e:fractal},
when $N$ is large
the huge majority of the $N/5$ eigenvalues of $B_{5,N,{\rm ev}}$ 
must be contained within an asymptotically small
neighbourhood of the origin; if $\cA _r$ intersects this neighbourhood, the law \eqref{e:fractal} 
necessarily fails, since the counting function
is proportional to $N$ instead of $N^\mu$. For the values of $N$ listed
in the table, this small region seems to be of radius $\gtrsim 0.005$, explaining the departure
from the fractal law in the last two columns.
%%%%%%%%%%%%%%%%%%%%%%%%%%%%%%
  \begin{figure}[htbp]
    \centerline{\includegraphics[width=13cm]{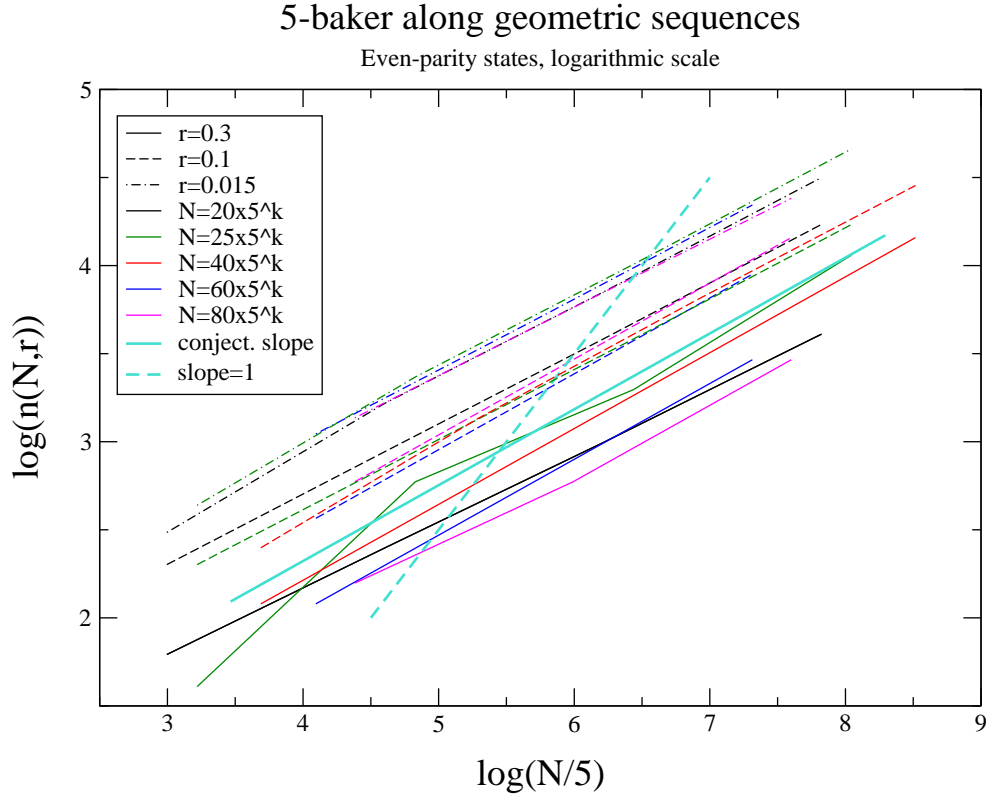}}
    \caption{\label{f:table3e} Checking the $N$-dependence of $n(N,r)$ for various
      values of $r$ and $N$ along geometric sequences $\{N_o\times 5^k\}$. We also show
(thick line) the conjectured slope $\log 2/\log 5$, and give for comparison the 
slope $1$ (thick dashed).}
  \end{figure}
%%%%%%%%%%%%%%%%%%%%%%%%%%%%%%
To further test the validity of the fractal law \eqref{e:fractal}, we choose a set of values of
$r$, and study the $N$-dependence of $n(N,r)$, for $N$ taken along several geometric
sequences, generalizing the above table.
In Fig.~\ref{f:table3e}, we plot this
dependence in logarithmic scale for $r=0.3$ (full lines), $r=0.1$ (dashed lines), 
$r=0.015$ (dot-dashed lines). 
Different geometric sequences are represented by a different colors. For almost all pairs $(N_o,r)$ 
the points
are almost aligned, and the slope is in very good agreement with the
conjectured one $\mu=\log 2/\log 5$. The less convincing data are the ones related to $r=0.3$:
for this radius, the numbers $n(N,r)$ are still quite small, so that fluctuations 
are much more visible than for the smaller radii. We expect this effect to 
disappear for larger values of $N$.

In Fig.~\ref{f:table3e} the {\em height} of the curves
does not only depend on $r$, 
but also on the sequence $\{N_o\times 5^k\}$ considered, especially for $r=0.3$.
%%%%%%%%%%%%%%%%%%%%%%%%%%%%%%%%%
\begin{figure}[htbp]
\centerline{\rotatebox{-90}{\includegraphics[width=20cm]{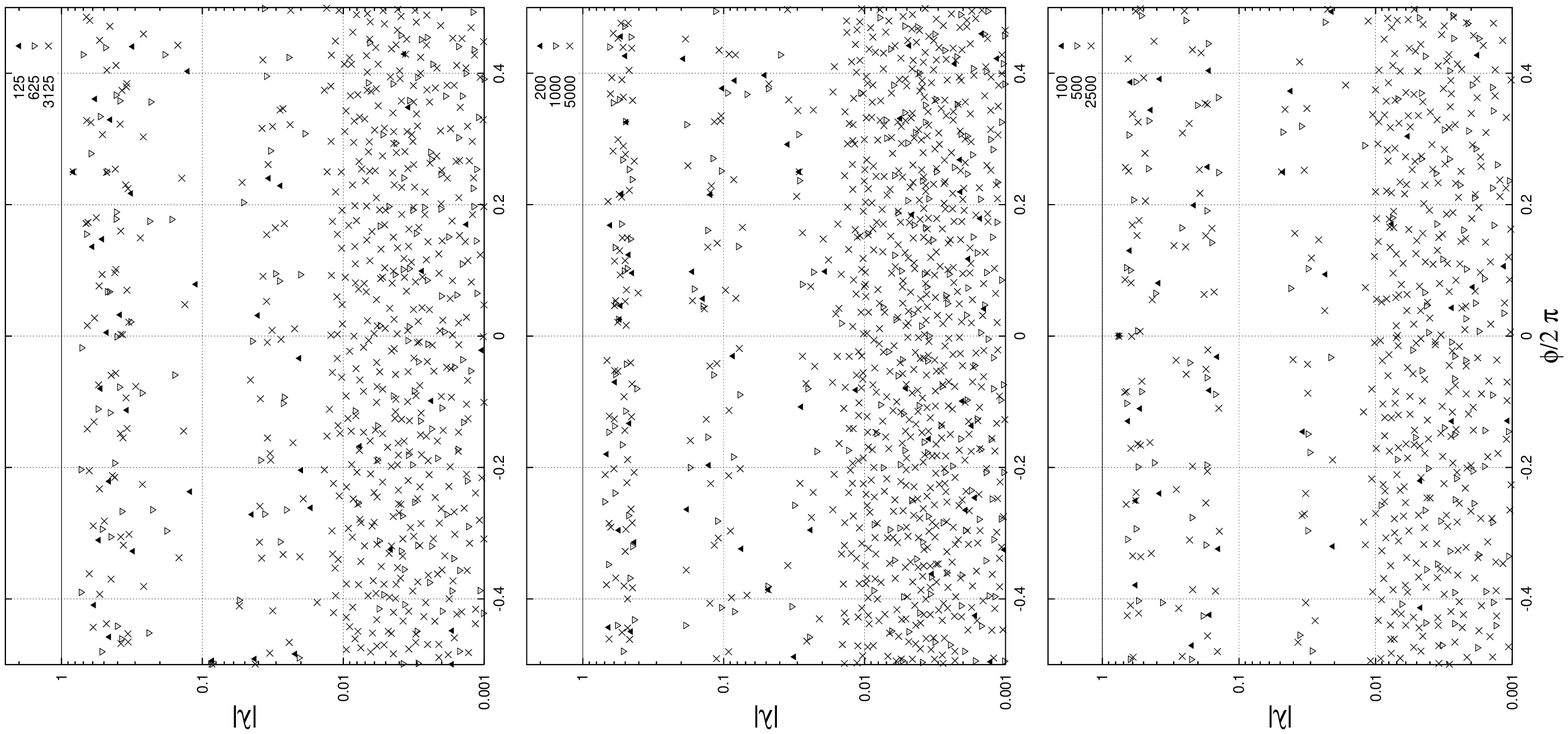}}}
\caption{\label{f:spect3be} Even-parity spectra of the  quantum baker's maps
$B_{5,N}$, along geometric sequences $N=N_o\times 5^k$. 
The eigenvalues are represented
on a logarithmic scale ($\arg\lambda/2\pi$ against $\log |\lambda|$). The dimensions
indicated correspond to $N/5$.}
\label{f:plat}
\end{figure}
%%%%%%%%%%%%%%%%%%%%%%%%%%%%%%%%%
 To investigate this 
apparent contradiction with \eqref{e:fractal}
 we plot in Fig.~\ref{f:spect3be} the even spectra of $B_{5,N}$ 
along 3 different geometric sequences.
These plots suggest that, along a given geometric sequence,
the eigenvalue density increases with $N$ uniformly with respect to 
 $\phi=\arg\lambda$, but very nonuniformly with respect to 
$|\lambda|$. We see that some
regions $\{r_0<|\lambda|<r_1\}$ remain empty even for large values of $N$. The presence of 
gaps was already noticeable when comparing the second and
third columns of table~\ref{table2}: obviously, for $N=20\times 5^k$, there were no
eigenvalues in the annulus $\{0.05<|\lambda|<0.1\}$, which is confirmed visually in
Fig.~\ref{f:spect3be} (bottom). This non-uniform dependence on $|\lambda|$ 
implies that the profile function $C(r)$ is nontrivial.

The spectra for the two other geometric sequences 
also shows the presence of gaps, but the gaps
differ from one geometric sequence to the other. This observation also
contradicts the law \eqref{e:fractal}.
%%%%%%%%%%%%%%%%%%%%%%%%%%%%%%
\begin{figure}[htbp]
  $$\includegraphics[height=11cm]{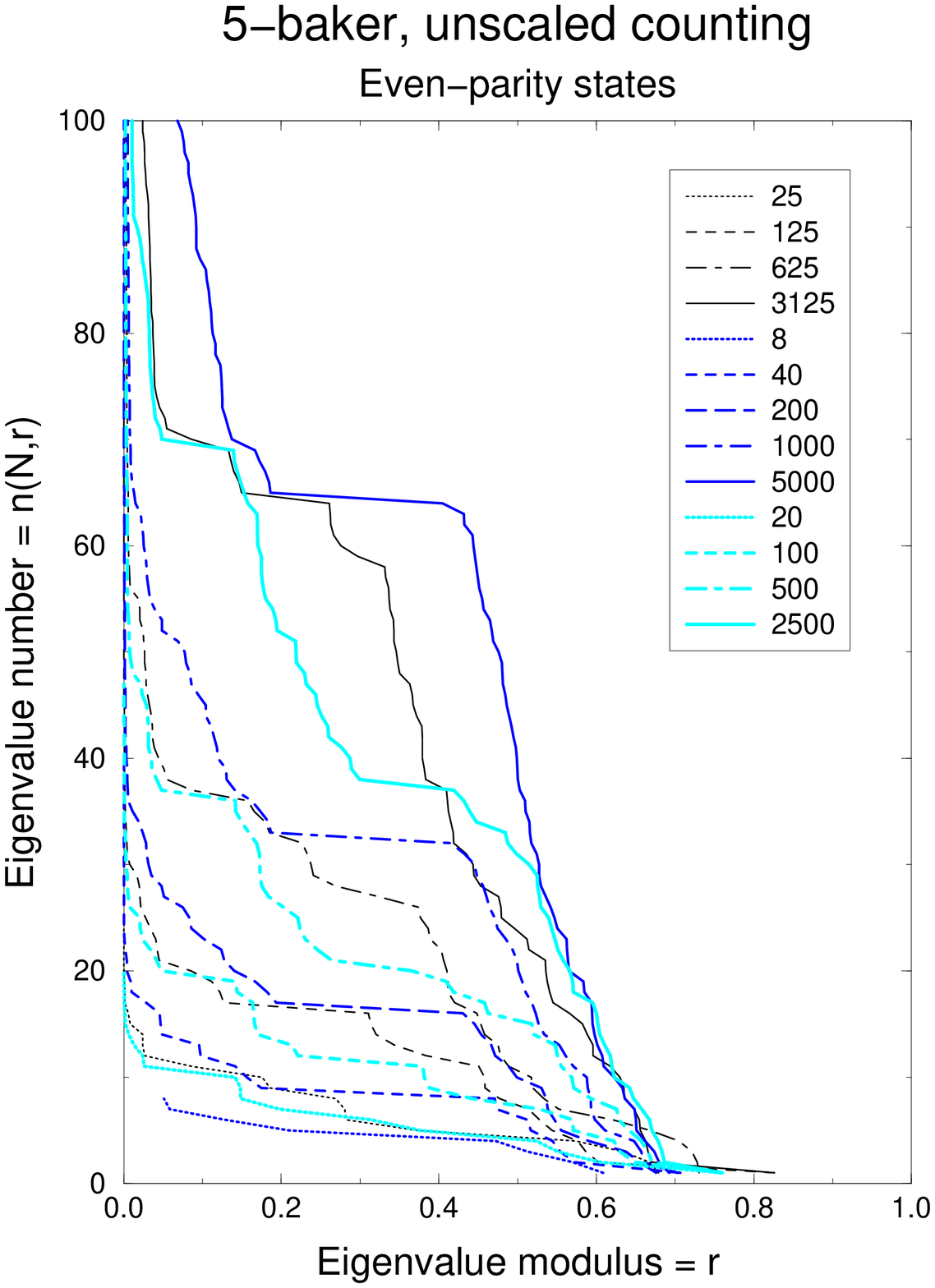}\quad
    \includegraphics[height=11cm]{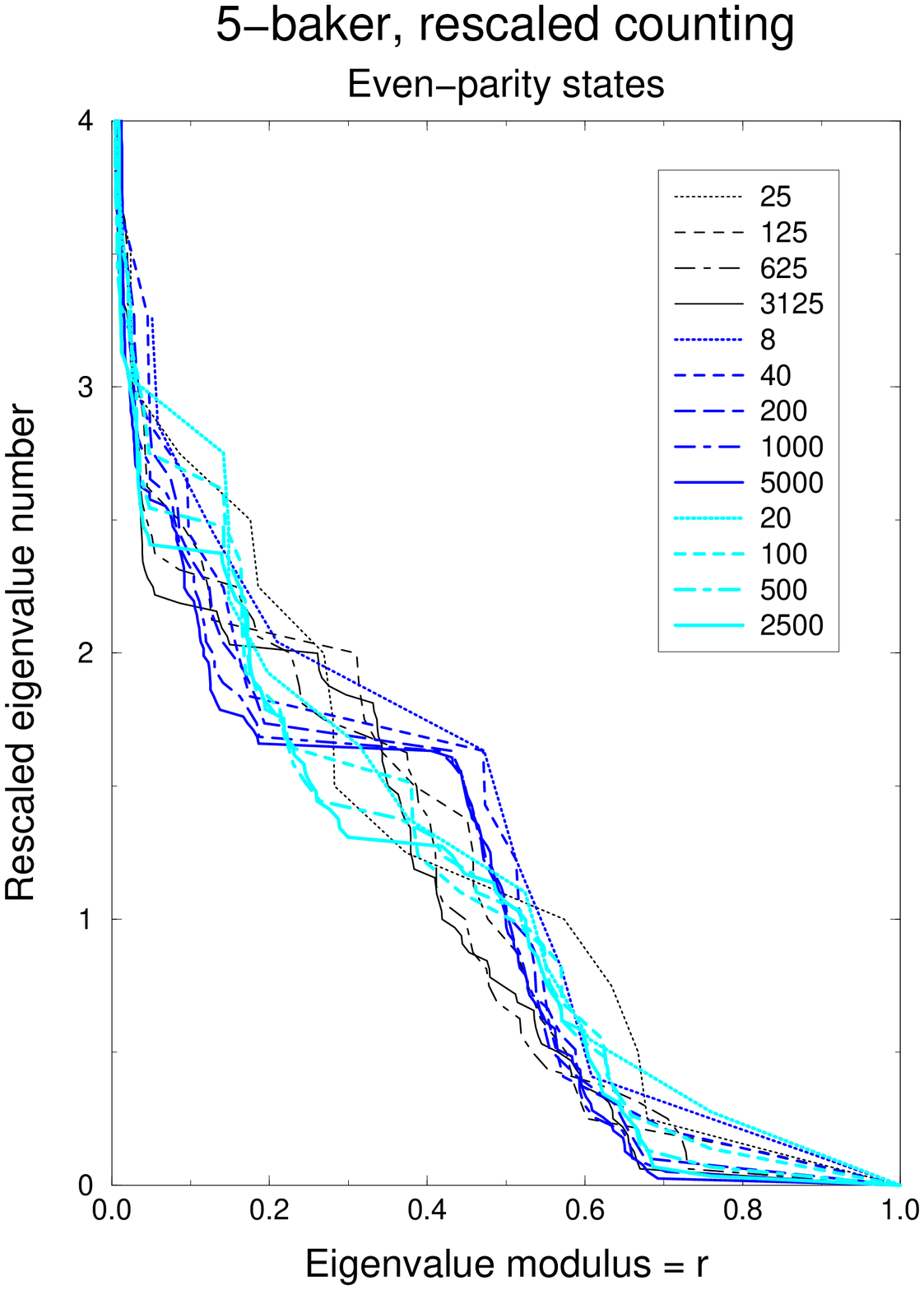}$$
    \caption{\label{f:spect-histo5be} On the left, we plot $n(N,r)$ as functions of  $r\in(0,1)$.
The numbers in the legend are $N/5$. On  the
right we have multiplied $n(N,r)$ by the factor $(N/5)^{-{\log 2}/{\log 5}}$.}
\end{figure}
%%%%%%%%%%%%%%%%%%%%%%%%%%%%%%
In spite of these problems we 
nevertheless attempt to compute the profile function $C(r)$ appearing 
in \eqref{e:fractal}.
Fig.~\ref{f:spect-histo5be} (left) shows 
$n(N,r)$ as functions of $r\in(0,1)$, for $N$ along the same three
geometric sequences (each one corresponding to a given color/width).
We then rescale the vertical coordinate of each curve
by the factor $(N/5)^{-{\log 2}/{\log 5}}$, and plot the rescaled curves in
Fig.~\ref{f:spect-histo5be} (right).
From far away, these rescaled curves are fairly superposed on each other, 
which shows that
the conjectured scaling \eqref{e:fractal} is approximately correct. Yet, 
a closer inspection shows that a much better convergence to a single function
occurs along each individual geometric sequence. For instance, the curves for 
$N=8\times 5^k$ ``pointwise'' converge to the last one along this sequence ($N/5=5000$),
which has a plateau on $\{0.2\lesssim r\lesssim 0.4\}$ corresponding to a spectral gap.
The curves of the two other sequences seem to converge as well, with plateaux on different 
intervals.

In the case of the open kicked rotator studied in \cite{schomerus}
the rescaled curves $n(N,r)$ are more or less superposed, therefore defining a
profile function $C(r)$. The authors claim that this function corresponds
reasonably well with a prediction 
of random matrix theory \cite{zycz}.
Our results for the 5-baker's map contradict this universality: 
there does not seem to be a global profile function $C(r)$, but a family of such functions,
which depend on the geometric sequence $\{N_o\times 5^k\}$, which could be denoted by $C(N_o,r)$.
The law \eqref{e:fractal} needs to be adapted by restricting $N$ to geometric sequences, which
yields the following empirical scaling law:

For any $N_o>0$ and $r\in(0,1)$, there exists $C(N_o,r)\geq 0$ such that, 
for $N$ along $\{N_o\times 5^k\}$ and $k\to\infty$,
\bequ\label{e:Weyl-sequ}
\# \set{ \lambda \in \Spec ( B_{N} )\cap \cA_{r,\vartheta,\rho}}  \simeq  
\ \frac{\rho}{2\pi}\, C(N_o,r) \,
N^{\mu } \,, 
\end{equation}
where $ \cA_{r,\vartheta,\rho} $ is given by \eqref{e:sector}.
In Fig.~\ref{f:spect-histo5be} 
the different profile functions are uniformly bounded,
$C_1(r)\leq C(N_o,r)\leq C_2(r)$, for some envelope functions $0\leq C_1(r)\leq C_2(r)$.

This weakening
 of \eqref{eq:weyl-1} to geometric sequences makes sense for baker's maps of 
the form $B_3$, $B_5$, which each have a uniform integer expansion factor, leading to number-theoretic 
properties.
In the case of a nonlinear open chaotic map (as the open kicked rotator of \cite{schomerus}), 
there is no reason for geometric sequences to play any role, so we expect \eqref{eq:weyl-1}
to hold in that case.

%%%%%%%%%%%%%%%%%%%%%%%%%%%%%%%%%%%%%%%%%%%%%%%%%%%%%%%% 
%%%%%%%%%%%%%%%%%%%%%%%%%%%%%%%%%%%%%%%%%%%%%%%%%%%%%%%% 
\section{A computable model}
\label{cm}
Because we are unable to analyze the spectra of the quantum bakers $B_N$ rigorously,
we introduce simplified models. In the case of the 3-baker,
we observe (see Fig.~\ref{fig:dens}, left) that the largest matrix elements 
are maximal along the ``tilted diagonals'' 
\bequ\label{e:diags}
(n,m)=(3l+\ep,l+\ell N/3),\quad \text{with}\quad l\in\{0,\ldots,N/3-1\},
\quad \ell\in\{0,2\},\quad\ep\in\{0,1,2\}\,.
\end{equation}
These ``diagonals''correspond to a discretization of the
the map $B_3$ projected on the position axis. Away from them, 
the coefficients do not decrease very fast due to the Gibbs phenomenon (diffraction).
The elements on the ``diagonals'' have moduli $1/\sqrt{3}+\Oo(1/N)$ and their phases only depend
on $\ell,\,\ep$ in the above parametrization.
Our simplified model is obtained by keeping only the elements on the ``diagonals'' 
(see Fig.~\ref{fig:dens}, right), set their moduli to $1/\sqrt{3}$
and shift their phases by $\pi/2$ (for convenience). Using the parametrization \eqref{e:diags},
we get:
\bequ
\label{eq:ttoy}
(\tB_{3,N})_{n\,m} =\frac{1}{\sqrt{3}}\,
\exp\big(\frac{2 i \pi}{3}(\ep+1/2)(\ell+1/2)\big) \,.
\end{equation}
For $N=9$ and using $\omega = e^{ 2 \pi i / 3 }$, the matrix reads
$$
\tB_{3,9} = \frac{\omega^{1/4}}{\sqrt 3} \left( \begin{array}{lllllllll}
      1 & 0 & 0 & 0 & 0 & 0 & \omega & 0 & 0 \\
      \omega^{1/2} & 0 & 0 & 0 & 0 & 0 & \omega^{1/2} & 0 & 0 \\
      \omega & 0 & 0 & 0 & 0 & 0 & 1 & 0 & 0 \\
      0 & 1 & 0 & 0 & 0 & 0 & 0 & \omega & 0 \\
      0 & \omega^{1/2} & 0 & 0 & 0 & 0 & 0 & \omega^{1/2} & 0 \\
      0 & \omega & 0 & 0 & 0 & 0 & 0 & 1 & 0 \\
      0 & 0 & 1 & 0 & 0 & 0 & 0 & 0 & \omega \\
      0 & 0 & \omega^{1/2} & 0 & 0 & 0 & 0 & 0 & \omega^{1/2} \\
      0 & 0 & \omega & 0 & 0 & 0 & 0 & 0 & 1
    \end{array} \right)\,.
$$
%%%%%%%%%%%%%%%%%%%%%%%%%%%
 \begin{figure}[htbp]
\centerline{\rotatebox{-90}{\includegraphics[height=15cm]{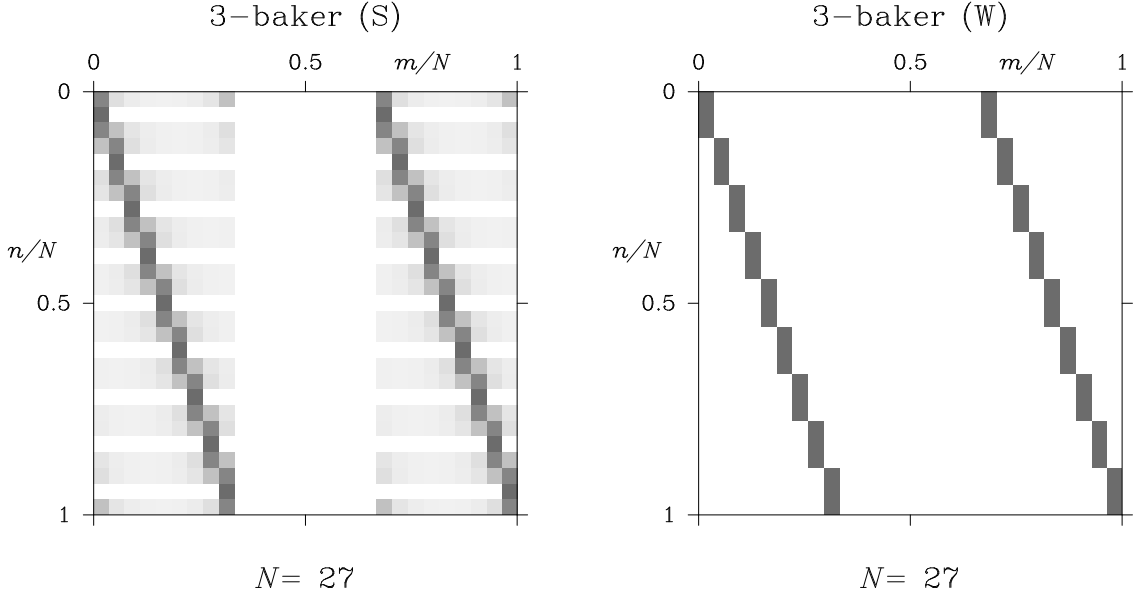}}}
\caption{Matrices  $B_{3,27}$ (left) and
its toy model $\tB_{3,27}$ (right). The gray scale represents the
modulus of the matrix elements (white$=0$, black$=1$).} 
\label{fig:dens}
\end{figure}
%%%%%%%%%%%%%%%%%%%%%%%%%%%%%%%
The matrix  $\tB_{3,N} $ can obviously not be considered as a ``small perturbation''
of $B_{3,N}$, since we removed many nonnegligible ``off-diagonal'' elements. Actually, 
by acting with $\tB_{3,N}$ on Gaussian coherent states, one realizes that these
matrices do not quantize the open 3-baker $B_3$ \eqref{eq:exb}, 
but rather a more complicated multivalued map $\tB_3$, built upon $B_3$
as follows:
\begin{equation}
\label{eq:exbt}
\forall (q,p)\in \cD_3,\quad\tB_3(q,p)  = \bigcup_{j=-1}^1 \{ B_3(q, p ) +(0,j/3)\} \,.
\end{equation}
We refer to \cite[Proposition 6.1]{NZ} for 
a precise statement. As opposed to $B_3$, the multivalued map \eqref{eq:exbt} is no longer 
obtained by truncating a canonical transformation, but it comes 
from three different transformations. $\tB_3$ can be considered
as a model of propagation with ray splitting. Another interpretation 
is given by considering a Markov process with 
probabilities $P(x',x)$ being allocated at each step 
to the image points $x'$ of $x$. 
Explicitly, the probabilities take the form 
$$
P(x',x)=f\left(\frac{3p'-[3q]-1/2}{3}\right) \,, \ \ 
f(t)=\left(\frac{\sin(3\pi t)}{3\sin(\pi t)}\right)^2 \,, \ \
x = ( q , p ) \,, \ x' = ( q' , p ') \,, 
$$ 
so that for each 
$x\in\cD_3$, the sum of the weights associated with the three images of $x$ is indeed $1$.

Some of the characteristics of the 
dynamics remain the same as for $B_3$. The local dynamics of each branch is the same, and
$\tB_3$ sends all points in $\t2\setminus \cD_3$ to infinity. 
One can define incoming and outgoing tails for $\tB_3$ (see \S\ref{obq}). 
As opposed to the case of $B_3$, these
tails are not symmetrical any more: 
$ \Gamma_- = \cC_3 \times [0,1)$, $\quad \Gamma_+ = \t2 $.
Yet, these formulas are slightly misleading. The second one comes from the fact that
any point $x\in \t2$ has two preimages through $\tB_3$, namely $x_0=(q/3,3p),\ x_2=((q+2)/3,3p)$,
so no point ever escapes to infinity in the past.
However, to these preimages are associated the respective weights $P(x,x_0)$, $P(x,x_2)$, the
sum of which is generally $<1$: there is thus a loss
of probability through $\tB_3^{-1}$, which is not accounted for by the definition of $\Gamma_+$.

In the next section we will show that the matrices $\tB_{3,N}$ can nonetheless be 
interpreted as quantizations of the original open baker $B_3$, as long as
one switches to a different notion of quantization, derived from a different type of
Fourier transform (the {\em Walsh-Fourier transform}).  

Families of {\em unitary} matrices $\tA_{2,N}$ with a structure similar to 
$\tB_{3,N}$  have already been proposed
as an alternative quantization of the 2-baker's map $A_2$ \cite{schack}.
These matrices can also closely related with the ``semiquantum bakers'' introduced in 
\cite{SaVo}\footnote{M.~Saraceno, private communication.}.
In the context
of quantum graphs (a recently popular model for quantum chaos), unitary matrices similar
with $\tA_{2,N}$ (but with random phases) occur 
as ``unitary transfer matrices'' associated with binary graphs \cite{tanner}. In this
framework, the matrix $\tA_{2,N}$ would correspond to a graph with very degenerate
bond lengths. 
In this framework, the matrix
$\tB_{3,N}$ is directly related with a classical transfer matrix defined by 
$(\cB_{3,N})_{j\,j'}=\big|(\tB_{3,N})_{j\,j'}\big|^2$, which represents the classical
Markov process on the graph. In our case, this transfer matrix is the discretized
version of the 
transfer (Perron-Frobenius) operator associated with the open map $B_3$.

%%%%%%%%%%%%%%%%%%%%%%%%%%%%%%%%%%%%%%%%%%%%%%%%%%%%%%%%%%%%%%%%%
\subsection{The Walsh model interpretation of $ \tB_{3,N}$.}
\label{wmi}
In this section, we represent the matrices $\tB_{3,N}$ in a way suitable for 
their spectral analysis. This can be done only in the case where $N$ is a power of
$3$. This representation is connected with the {\em Walsh model} of harmonic analysis.

The latter originally appeared in the context of fast signal processing \cite{lifermann}.
The major advantage of Walsh harmonic analysis (compared with the usual Fourier analysis)
is the possibility to strictly localize  wave packets simultaneously
in position and in momentum. For our
problem, this has the effect to remove the diffraction problems
due to the discontinuities of the classical map, which spoil the usual semiclassics \cite{SaVo}.

A recent preprint \cite{arul} analyzes some special eigenstates
of the ``standard'' quantum 2-baker,
using the Walsh-Hadamard transform (which slightly differs from the
Walsh transform we give below) as a ``filter''. We are doing something
different here by constructing our simplified model $\tB_{3,N}$ from the
Walsh transform, as $B_{3,N}$ was constructed from the discrete Fourier transform (see \S\ref{obq}).

We first select the expanding
coefficient of the baker's map, which we denote by $D\in\NN$
(the map \eqref{eq:exb} is associated with $D = 3$, the map \eqref{eq:exA} with $D=5$).
Once this is done, we will restrict ourselves to the values of $N$ along
the geometric sequence $\{N=D^k\,,k\in\NN\}$. In this
case, the Hilbert space can be naturally decomposed as a tensor product of $k$ spaces
$\CC^D$:
\bequ\label{e:tensor}
\hn= (\CC^D)_1\otimes(\CC^D)_2\otimes\cdots\otimes(\CC^D)_k\,.
\end{equation}
This decomposition appears naturally in the context of quantum computation, where
each $\CC^D$ represents a ``quantum $D$-git'', that is, a quantum system with 
$D$ levels. Here, we realize this decomposition using the basis
of position eigenstates $|q_j\ra$ of $\hn$ (see \cite{schack} for the case $D=2$).
Indeed, each quantum position $q_j=({j+1/2})/{N}$, $j\in\ZZ_{D^k}=\{0,\ldots,N-1\}$ 
is in one-to-one correspondence with
a {\sl word} $\bep=\ep_1\ep_2\cdots \ep_k$ made of {\sl symbols} ($D$-gits) 
$\ep_\ell=\ep_\ell(j)\in \ZZ_{D}$:
\begin{equation}\label{e:Dit}
j=\sum_{\ell=1}^k \ep_\ell\,D^{k-\ell}\,.
\end{equation}
The usual order for $j\in\ZZ_{D^k}$
corresponds to the lexicographic order for the symbolic words $\bep\in(\ZZ_D)^k$.
Associating to each $D$-git a $D$-dimensional vector space $(\CC^D)_{\ell}$
with canonical basis $\{e_0,e_1,\ldots, e_{D-1}\}$, the position eigenstate $|q_j\ra\in\hn$
can be decomposed as
\begin{equation}\label{e:quDit}
|q_j\ra= e_{\ep_1}\otimes e_{\ep_2}\otimes\cdots \otimes e_{\ep_k}\,.
\end{equation}
This identification realizes the tensor product decomposition \eqref{e:tensor}.

The Fourier transforms $\G_N$ \eqref{e:Fourier1}, and the simpler one without 
the $1/2$ shift,
\bequ\label{e:Fourier2}
(\F_N)_{j\,j'}=\frac{e^{-2 i \pi  j\,j'/N}}{\sqrt{N}}\,,\quad j,j'\in\ZZ_N\,,\ N=D^k\,,
\end{equation}
are defined by applying the exponential function $x\mapsto e^{-2 i \pi x}$ to the products
$$
\frac{jj'}{D^k}=\sum_{m=2-k}^{k} D^{-m}\,\tilde\ep_m(jj'),\quad
\text{where}\quad\tilde\ep_m(jj')=\sum_{\ell+\ell'=m+k}\ep_\ell(j)\ep_{\ell'}(j')\,.
$$
If we replace in \eqref{e:Fourier2} the exponential 
$e^{-2 i \pi x}$ by the piecewise constant function
$e_D(x)=\exp(-2 i \pi [Dx]/D)$, and replace each $\tilde\ep_m(jj')$ by its value 
$\ep_m(jj')$ modulo $D$, we obtain the matrix element 
\begin{equation}\label{e:walshV}
\begin{split}
(\cV_{k})_{j\,j'}&\defeq D^{-k/2}\,e_D\big(\sum_{m=2-k}^{k} D^{-m}\ep_m(jj')\big)
=D^{-k/2}\,\exp\big(-\frac{2 i \pi}{D}\, \ep_1(jj')\big)\\
&=\prod_{\ell=1}^k D^{-1/2}\,\exp\Big(-\frac{2 i \pi}{D} \,\ep_\ell(j)\,\ep_{k+1-\ell}(j')\Big)\,.
\end{split}
\end{equation}
The matrix $\cV_{k}$ defines the Walsh transform in dimension $D^k$. 

Because we have used the ``half-integer'' Fourier transform \eqref{e:Fourier1} 
to define our quantum baker's map, we will need a slightly different version of
Walsh transform, namely 
$$
(\cW_{k})_{j\,j'}\defeq \prod_{\ell=1}^k D^{-1/2}\,
\exp\Big(-\frac{2 i \pi}{D}\, \big(\ep_\ell(j)+1/2)\,(\ep_{k+1-\ell}(j')+1/2\big)\Big)\,.
$$
Both $\cV_k$ and $\cW_k$
preserve the tensor product structure \eqref{e:tensor}: 
for any $v_1,\ldots,v_n\in\CC^D$,
\bequ\label{e:actionW}
\cV_k ( v_1 \otimes \cdots \otimes v_k ) 
=  \F_D v_k \otimes \cdots \otimes \F_D v_1  \,, \qquad \cW_k ( v_1 \otimes \cdots \otimes v_k ) 
=  \G_D v_k \otimes \cdots \otimes \G_D v_1  \,.
\end{equation}
These expressions show
that $\cV_k$ and $\cW_k$ are {\sl unitary}.

Specializing the computations to $D=3$, we are now in position to 
define the toy model $ \tB_{3,N} $ (in the case $N=3^k$) as the ``Walsh-quantization'' of the
3-baker \eqref{eq:exb} (as opposed to the ``standard'' quantization of the multivalued
map $\tB_3$ \eqref{eq:exbt}). Indeed, one can check that the matrix \eqref{eq:ttoy} 
can be expressed as
\begin{equation}
\label{eq:BNc}
\tB_{3,N} = 
\cW_k^* \left( \begin{array}{lll} \cW_{k-1} &  0 & \ 0 \\
\ 0 &  0  &  \ 0 \\
\ 0 &  0 & \cW_{k-1} \end{array} \right) \,.
\end{equation}
This formula is clearly the Walsh analogue of the definition \eqref{eq:BN} of the ``standard''
quantum open baker $B_{3,N}$. From this definition and \eqref{e:actionW}, 
we see the action of  $B_{3,N}$ on tensor products:
\bequ\label{e:decompo-B}
\tB_{3,N}( v_1 \otimes \cdots \otimes v_k )=
v_2 \otimes \cdots \otimes v_k\otimes \G_3^*\pi_{0,2} v_1\,,
\quad  v_j\in\CC^3 \,,
\end{equation}
where $\pi_{0,2}$ is the orthogonal projector (in $\CC^3$) on $\CC e_0 \oplus \CC e_2 $. 

%%%%%%%%%%%%%%%%%%%%%%%%%%%%%%%%%%%%%%%
\subsection{Distribution of resonances}
\label{dr}

%%%%%%%%%%%%%%%%%%%%%%%%%%%%%
\begin{figure}[htbp]
\begin{center}
\rotatebox{-90}{\includegraphics[height=10cm]{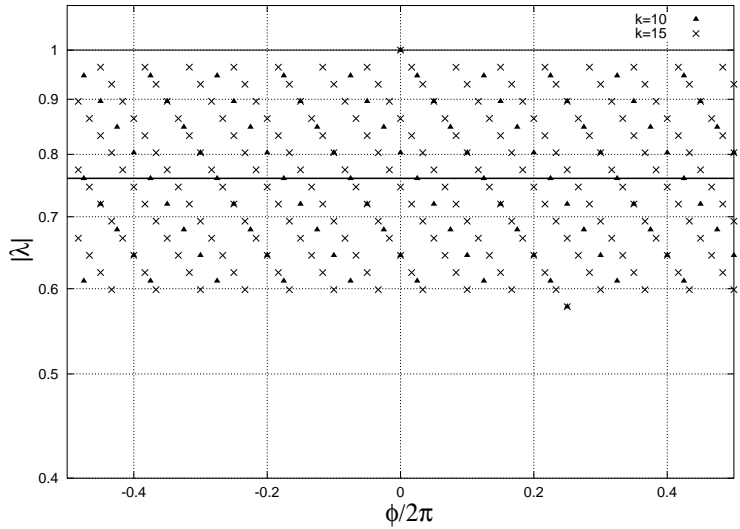}}
\caption{
Eigenvalues of the matrices $ \tB_{3,N}$
for $ N =  3^{10} $ (triangles) and for $ N = 
 3^{15} $ (crosses), forming lattices in a logarithmic scale. 
The two horizontal lines correspond to the spectral radius 
$ |z|=1$ (thin line)  and the ``peak multiplicity'' $|z|= 3^{-1/4}$ (thick line). 
Notice the difference of vertical scale
compared with the spectra of Fig.~\ref{f:spect3be}.}
\label{f:lat}
\end{center}
\end{figure}
%%%%%%%%%%%%%%%%%%%%%%%%%%%%%
Using \eqref{e:decompo-B} we can explicitly describe the spectrum of $ \tB_{3,N} $ for $N=3^k$. 
The computation is identical with \cite[Section~6.2]{NZ}, so we only give the results.
The generalized kernel of $\tB_{3,N}$ is spanned by the position states $|q_j\ra$ such that $\ep_\ell(j)=1$ 
for at least one index $1\leq\ell\leq k$. 
This corresponds to positions $q_j$ ``far'' from the Cantor set $\cC_3$, so that 
the classical points $(q_j,p)$ are
sent to infinity at a time $n\leq k$. This kernel has dimension $3^k-2^k=N-N^{\log 2/\log 3}$.

The nonzero eigenvalues of $\tB_{3,N}$ are given by the set (see Fig.~\ref{f:lat})
\[
\{ \lambda_+ \} \cup \{ \lambda_- \} \cup
\bigcup_{\ell=0}^{k-1}\bigcup_{p=1}^{k-1} \{ e^{2 i \pi \ell/k}\, 
\lambda_+^{1-p/k}\,\lambda_-^{p/k}\} \,,\quad\text{where}\ 
\lambda_+=1,\ \lambda_-=\frac{i}{\sqrt{3}}\,.
\]
For each $p\in\{1,\ldots,k-1\}$, the $k$ eigenvalues of modulus $|\lambda_-|^{p/k}=3^{-p/2k}$ 
asymptotically have the same
degeneracy $\binom{k}{p}/k $ as $k\to\infty$ (semiclassical limit), 
which shows that their distribution is
circular-symmetric. Taking these multiplicities into account, we obtain the following
Weyl law for the eigenvalues of $\tB_{3,N}$ inside a region \eqref{e:sector},
along the sequence $N\in\{3^k\}$, $k\to\infty$:
\begin{gather}
\label{eq:weyl}
\begin{gathered}
 \#\set{ \Spec ( \tB_{3,N} ) \cap \cA_{r,\vartheta,\rho} }  \
= \frac{\rho}{2\pi}\,\,N^{\mu }\, ( C(1,r) + o(1))  
\,  \\ 
\mu = \dim ( \Gamma_- \cap W_{+} ) = \frac{\log2}{\log3} \,, \qquad
C (1, r ) = \bbbone_{ ( 0, 3^{-1/4} ] } ( r ) \,.  
\end{gathered}
\end{gather}
The values $\lambda_-$, $\lambda_+$ in \eqref{eq:weyl} are the nonzero eigenvalues of the matrix 
$\G_3^*\circ\pi_{0,2}$ appearing in \eqref{e:decompo-B}. We used the notation $C(1,r)$ for the
profile function to be consistent with our notations in \eqref{e:Weyl-sequ}, that is, to emphasize
that this estimate is valid only along the sequence $N\in\set{1\times 3^k}$.

We notice that the spectrum of the classical transfer
matrix $\cB_{3,N}$ defined at the end of \S\ref{cm} is drastically different: 
this matrix admits one simple nontrivial
eigenvalue $\lambda=2/3$ (interpreted as the classical escape rate), the rest of the
spectrum lying in the generalized kernel. Therefore, the features of the quantum spectrum is
intimately related with the oscillatory phases of $\tB_{3,N}$ along the ``diagonals''.

%%%%%%%%%%%%%%%%%%%%%%%%%%%%%%%%%%%%%%%
\subsection{Conductance and Shot Noise}
\label{ff}

In this section, we consider
 an open baker's map as a model of quantum transport
through a ``chaotic quantum dot'', that is a 2-dimensional cavity connected to the
outside world through a certain number of ``leads'' carrying the current; each lead is
connected to the cavity along a segment $L_j$ of the boundary (see Fig.~\ref{f:dot}), and the
connection is assumed to be ``perfect'': a particule inside the cavity which 
hits the boundary along $q\in L_j$
is completely evacuated to the lead. Therefore, the phase space domain
$L_j\times [0,1)$ above this segment is a part of the ``hole'', in the terminology of \S\ref{obq},
whereas the remaining set $I=[0,1)\setminus(\cup L_j)$ 
represents the boundary of the quantum dot, which lifts to the phase space
domain $\cD=I\times [0,1)$.

In the previous sections we have studied the open quantum map obtained by 
projecting a unitary quantum dynamics (called generically $U_N$) onto a subdomain
$\cD$ of the phase space: resonances were defined as the eigenvalues of 
$U_N\Pi_{\cD}$. These resonances are supposed to represent the metastable quantum
states inside the open quantum dot, after it has been opened.
In the present section, we want to study another aspect of the open system, namely 
the ``transport'' through the dot, using the formalism
of \cite{beenakker}. We will focus on the case where the opening $L$ splits into 
two segments $L=L_1\cup L_2$, and we study the transmission matrix 
from the lead $L_1$ to the lead $L_2$.
%%%%%%%%%%%%%%%%%%%%%%%%%%%%%%%%%%%
\begin{figure}[htbp]
\begin{center}
\includegraphics[width=5.5in]{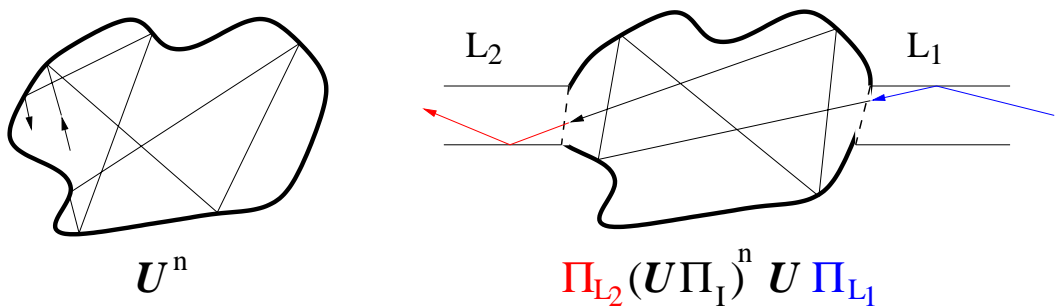}
\caption
{\label{f:dot}
Closed and open chaotic cavity. On the left, a (bounded) trajectory
is schematically associated with a power of the unitary quantum propgator ($n$ represents
the number of bounces). On the right, a ``transmitting'' trajectory is associated with a term
of the matrix \eqref{eq:def.t}.}
\end{center}
\end{figure}
%%%%%%%%%%%%%%%%%%%%%%%%%%%%%%%%%%%
Once we are given, on one side, the quantum map $U_N$ associated with the
{\sl closed} dynamics inside the ``cavity'',
on the other side, the projectors on the leads 
$\Pi_{L_i}$ and on the ``interior'' $\Pi_I=\Pi_\cD$,
the transmission matrix (from $L_1$ to $L_2$) is defined as the block
\begin{equation}
\label{eq:def.t}
t(\vartheta)=
\sum_{n\geq 1} e^{in\vartheta}\,\Pi_{L_2}\,U_N\,(\Pi_I U_N)^{n-1}\,\Pi_{L_1}\,.
\end{equation}
The parameter $\vartheta\in[0,2\pi)$ is the ``quasi-energy'' of the particles.
According to Landauer's theory of coherent transport, 
each eigenvalue $T_i(\vartheta)$ of the matrix 
$ t(\vartheta)t^*(\vartheta)$ corresponds to a ``transmission channel''. The 
dimensionless conductance of the system is then given by
\begin{equation}
\label{eq:cond}
g(\vartheta)=\tr\big(t(\vartheta)\,t^* (\vartheta)\big)\,.
\end{equation}
A transmission channel is ``classical'' if the eigenvalue $T_i$ is very close to
unity (perfect transmission) or close to zero (perfect reflection). The
intermediate values characterize ``nonclassical channels'' 
(governed by strong interference effects). The number number of the latter
can be estimated by the noise power
\bequ\label{eq:shot}
P(\vartheta)=\tr\Big( t(\vartheta)t^*(\vartheta)\big(Id- t(\vartheta)t^*(\vartheta)\big)\Big)\,,
\end{equation}
or equivalently the Fano factor, $F={P}/{g}$.  It is sometimes necessary to
perform an ensemble averaging over $\vartheta$ 
to obtain significant results \cite{beenakker}. However, for the model
we study here, these quantities will depend very little on $\vartheta$.
%%%%%%%%%%%%%%%%%%%%%%%%%%%%%%%%%%%
\begin{figure}[htbp]
\begin{center}
\includegraphics[width=5.0in]{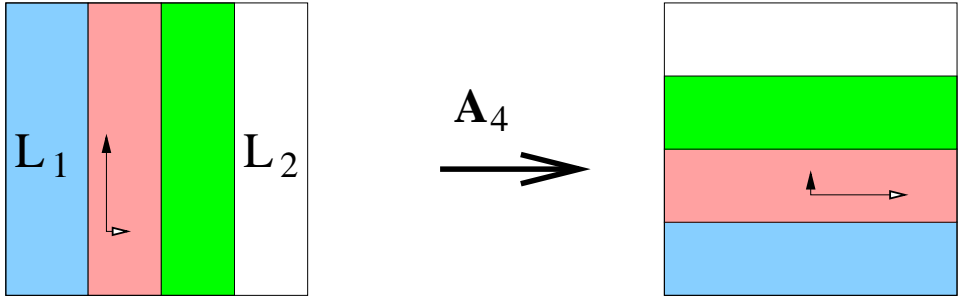}
\caption
{\label{f:4-baker}
4-baker's map modeling the chaotic cavity. The left- and rightmost vertical strips
correspond to the two openings (``leads'').}
\end{center}
\end{figure}
%%%%%%%%%%%%%%%%%%%%%%%%%%%%%%%%%%%
The closed quantum dot will be modeled by the following quantum map: 
we consider the 4-baker's map $A_4$ and quantize it using the  Walsh
transform $\cV_k$ \eqref{e:walshV} with $D=4$.
In dimension $N=4^k$, our unitary propagator is therefore
\[ 
U_N = \tA_{4,N}=
\cV_k^* \left( \begin{array}{llll} 
\cV_{k-1} &0&0&0 \\
0&\cV_{k-1} &0&0\\
0&0&\cV_{k-1}&0\\ 
0&0&0&\cV_{k-1} 
\end{array} \right) \,.
\]
We attach the
leads on the intervals $L_1=[0,1/4]$ and $L_2=[3/4,1]$: this way, the projectors 
$\Pi_{L_i}$ as well as the projector $\Pi_I=Id-\Pi_{L_1}-\Pi_{L_2}$ can be represented 
as tensor product operators:
$$
\Pi_{L_1}=\pi_{0}\otimes Id_4\otimes Id_4\otimes\cdots\,,\quad
\Pi_{L_2}=\pi_{3}\otimes Id_4\otimes\cdots\,,\quad
\Pi_I=\pi_{I}\otimes Id_4\otimes\cdots\,.
$$
Here $\pi_i$ is the orthogonal projector on the basis state $e_i$ of $\CC^4$,
and $\pi_I=\pi_1\oplus\pi_2$. This tensor action, together with the action of $\tA_{4,N}$
(analogous to \eqref{e:decompo-B}), allow us to compute all quantities explicitly.

The spectrum of the ``inside'' propagator 
for this model, $\tB_{4,N}=\tA_{4,N} \Pi_I$, 
satisfies a fractal Weyl law of the type \eqref{eq:weyl} along the sequence $ N = 4^k$,
with exponent $\mu = \log 2 / \log 4 = 1/2 $, 
and profile $ C (1,r) = \bbbone_{ [0, 2^{-3/4} ] } ( r ) $.

For this model and in the semiclassical limit $k\to\infty$, 
we could compute the dimensionless conductance \eqref{eq:cond}.
The computation \cite[\S 7.2]{NZ} requires to control
the time evolution up to $n=Ck$ for some $1<C<2$: this is of the
order of the Ehrenfest time $\tau_E=k$ for the system. For any 
$ \vartheta $ we obtain
\begin{equation}\label{eq:conda}
g ( N=4^k , \vartheta) = \frac{4^{k-1} }{2} ( 1 + o (2^{-k})) =
\frac{N/4}{2}\, ( 1 + o( 1 ) ) \,,\qquad k\to\infty\,.
\end{equation}
Here $ N/4 $ is 
the number of transmission channels from $L_1$ to $L_2$, that is the rank of 
the matrix $t(\vartheta)$.
We see that, as could be expected,
approximately one half of the scattering channels get transmitted
from one lead to the other, the other half being reflected back.

Asymptotics for the shot noise \eqref{eq:shot} (which counts the 
``nonclassical'' transmission channels'') are more interesting
and again independent of $ \vartheta$:
\bequ
\label{eq:shota}
P ( N=4^k,\vartheta ) = 
2^{k-1}\,\Big(\frac{11}{80}+\Oo(e^{-C k})\Big) = 
\frac{11}{80}\, (N/4)^{\mu }\, ( 1 + o(1) ) \,,\quad k\to\infty\,.
\end{equation}
Here $ \mu = 1/2$ is the dimension appearing in the 
fractal Weyl law for the resonances. A similar fractal law for the shot noise
had been observed in \cite{beenakker} in the case of the quantum
kicked rotator; the  power law $N^{\mu}$ for the number of nonclassical channels
was explained there through a study of the dynamics
up to the Ehrenfest time. 

The constant  $11/80$ in \eqref{eq:shota} gives
the average ``shot noise'' per nonclassical transmission channel.
This number is close to the random matrix theory prediction for
this quantity, namely $1/8$ \cite{jalabert,beenakker}. The precise number
$11/80$ certainly depends on which baker's map one starts from, and
which quantization one uses. For instance, we did not check whether the
``half-integer'' Walsh quantization of the 4-baker leads to the same
prefactor, but we expect the result to be close to it. It would be interesting
to actually check the full distribution for the transmission eigenvalues $T_i$, and
compare it with the prediction of random matrix theory \cite{jalabert}.
 
The near agreement with random matrix theory is in contrast with
the fact that the semiclassical resonance spectrum of the 
propagator $\tB_{4,N}$ inside the dot is very different from
that of a random subunitary matrix. Somehow, the matrix $t(\vartheta)$, obtained by summing iterates
of $\tB_{4,N}$, has acquired some ``randomness'', as far as the distribution of 
its singular values is concerned.

The transport properties of chaotic cavities has also been studied within the 
framework
of quantum graphs. The shot noise \eqref{eq:shot} could be semiclassically 
estimated in the case of a ``star graph'', by summing over
transmitting trajectories on the graph \cite{schanz} (they studied the case of ``small openings''). 
The authors show that one needs to take into account the ``action correlations'' 
between different trajectories, in order to reproduce the random matrix result.
As mentioned before, the matrix $\tA_{4,N}$ can be interpreted as the
unitary transfer matrix for a different type of graph \cite{tanner}, 
with bonds having degenerate lengths. Somehow, our use of the tensor product structure
implicitly takes into account the action correlations for this particular
graph.

%%%%%%%%%%%%%%%%%%%%%%%%%%%%%%%%%%%%%%%%%%%%%%%
\section{Conclusions}
\label{con}

Quantum open baker's maps provide a simple and elegant model for 
the study of quantum resonances of open chaotic systems. The numerical 
investigation of these models is easily accessible
and, as shown in \S\ref{nr}, gives a good agreement with the fractal Weyl law on 
``small energy scales'', which is \eqref{eq:fractal-hamil} in the case of 
Hamiltonian flows.  Only larger energy scales
\eqref{eq:fractal-large} were considered previously. It would be interesting to 
investigate the spectrum of the model operator \eqref{eq:BN} for
higher values of $N\sim\hbar^{-1}$. The na\"{\i}ve numerical approach we took 
(full diagonalization
of the matrices $B_N$) only allowed to reach values $ N \lesssim 5000 $. It would
make more sense to use an algorithm allowing us to extract only the largest
eigenvalues (which are the ones we are interested in), instead of the full spectrum.

By modifying the standard quantum baker's map, in a way which 
still fits in the framework of quantization of chaotic dynamics, 
we obtained a model for which the fractal Weyl law 
\eqref{eq:weyl} can be rigorously proven. Since the spectrum of this model is 
explicitely computable (and forms a lattice), it is 
forcibly nongeneric. However, the explicit computation of other physical quantities
associated with our model, namely the conductance and the ``shot noise'',
shows more generic properties. The fractal Weyl law is also present
in the calculation of the ``shot noise'', and the prefactor is (unexpectedly) close to
random matrix predictions.

\end{document}